\documentclass[conference]{IEEEtran}
\IEEEoverridecommandlockouts
\usepackage{cite}
\usepackage{amsmath,amssymb,amsfonts}
\usepackage{algorithmic}
\usepackage{graphicx}
\usepackage{textcomp}
\usepackage{xcolor}
\usepackage{subcaption}
\def\BibTeX{{\rm B\kern-.05em{\sc i\kern-.025em b}\kern-.08em
    T\kern-.1667em\lower.7ex\hbox{E}\kern-.125emX}}
\usepackage[inline]{enumitem}

\begin{document}

\title{Power Constrained Autotuning using Graph Neural Networks}

\author{\IEEEauthorblockN{Akash Dutta}
\IEEEauthorblockA{\textit{Iowa State University}\\
Iowa, USA \\
adutta@iastate.edu}
\and
\IEEEauthorblockN{Jee Choi}
\IEEEauthorblockA{
\textit{University of Oregon}\\
Oregon, USA \\
jeec@uoregon.edu}
\and
\IEEEauthorblockN{Ali Jannesari}
\IEEEauthorblockA{
\textit{Iowa State University}\\
Iowa, USA \\
jannesari@iastate.edu}
}

\maketitle
\thispagestyle{plain}
\pagestyle{plain}
\begin{abstract}
Recent advances in multi and many-core processors have led to significant improvements in the performance of scientific computing applications.
However, the addition of a large number of complex cores have also increased the overall power consumption, and power has become a first-order design constraint in modern processors.
While we can limit power consumption by simply applying power constraints,  applying them blindly will lead to non-trivial performance degradation.
To address the challenge of improving the performance, power, and energy efficiency of scientific applications on modern multi-core processors, we propose a 
novel Graph Neural Network based auto-tuning approach that 
\begin{enumerate*}[label=(\roman*)]
    \item optimizes runtime performance at pre-defined power constraints, and
    \item simultaneously optimizes for runtime performance and energy efficiency by minimizing the energy-delay product
\end{enumerate*}.
The key idea behind this approach lies in modeling parallel code regions as flow-aware code graphs to capture both semantic and structural code features. 
We demonstrate the efficacy of our approach by conducting an extensive evaluation on $30$ benchmarks and proxy-/mini-applications with $68$ {\tt OpenMP} code regions.
Our approach identifies {\tt OpenMP} configurations at different power constraints that yield a geometric mean performance improvement of more than $25\%$ and $13\%$ over the default {\tt OpenMP} configuration on a 32-core Skylake and a $16$-core Haswell processors, respectively.
In addition, when we optimize for the energy-delay product, our auto-tuner-selected {\tt OpenMP} configurations demonstrate both performance improvement of $21\%$ and $11\%$ and energy reduction of $29\%$ and $18\%$ over the default {\tt OpenMP} configuration at Thermal Design Power for the same Skylake and Haswell processors, respectively.
\end{abstract}

\begin{IEEEkeywords}
Auto-tuning, OpenMP, GNN, Power constraint
\end{IEEEkeywords}

\section{Introduction}
\label{sec:introduction}
High-performance computing (HPC) systems
have exploded in both capacity and complexity over the past decade, and 
this has led to substantial improvement in performance of various scientific applications.
However, more complex larger systems consume more power, and in the absence of expensive cooling solutions, increased power consumption leads to higher operational temperature and inefficient resource utilization (via higher static power, shorter device lifespan, and more).
As a result, power consumption has become a first-order hardware design constraint for modern multi- and many-core systems.
Unfortunately, focusing on hardware advancements for reducing power consumption is insufficient, as inefficient usage of the underlying hardware due to poor parallel coding practices may negate any hardware improvements.

Many software solutions currently exist for controlling power.
At the processor level, vendor-provided tools can be used to artificially lower power consumption.
For example, power consumption can be controlled in recent Intel processors using the Running Average Power Limit (RAPL) interface \cite{david2010rapl}, which ensures that an application does not exceed a predefined power budget.
However, a common drawback of a fixed power budget is that it \emph{slows down} execution by lowering the processor clock, and this can have adverse effects on real-time or time-bound applications.
At the data-center level, a common approach to reducing power consumption is through over-provisioning (i.e., have more hardware available than can be powered simultaneously at any time) and constraining the power limit for each node~\cite{10.1145/2464996.2465009}. 
In such a setting, a static algorithm for distributing power across nodes may lead to \emph{degraded throughput}, and a more sophisticated approach that adjusts the execution dynamically is required to harness the full potential of the underlying system.

One strategy to address both scenarios is to adjust the execution of the application directly, such that they meet some user-specified (e.g., individuals or data-centers) performance and/or power constraints.
This will allow users to tailor their application to domain-specific environments (e.g., edge or mobile computing) or design scheduling policies for data-center power management.
{\tt OpenMP}, as the de-facto parallel programming model for intra-node parallelism, provides a number of tunable parameters that highly influence code execution, which makes it highly suitable for this purpose.
While there is already a large body of work targeting performance tuning, there are only a few studies that target
power.
In addition, due to the large configuration search space for {\tt OpenMP} on modern multi- and many-core processors, most of these studies require multiple executions to determine the optimal configuration~\cite{tapus2002active, ansel2014opentuner, thiagarajan2018bootstrapping, roy2021bliss,bari2016arcs,bari2019performance}, which is both time consuming and resource intensive.


As a motivating example, we consider the \textit{ApplyAccelerationBoundaryConditionsForNodes} kernel from the {\tt LULESH} \cite{karlin2013lulesh} proxy application. 
On a $16$-core dual-socket Haswell processor with a Thermal Design Power (TDP) of 85W, an exhaustive search of the {\tt OpenMP} configuration space yields the highest speedups of $7.54\times$, $2.11\times$, $1.80\times$ and $1.67\times$ over the typical (or default) {\tt OpenMP} configuration at power constraints of $40$W, $60$W, $70$W and $85$W, respectively.
%
However, \emph{none of these {\tt OpenMP} configurations lead to the highest energy efficiency}.
The most energy-efficient execution occurs at a power constraint of 60W using a {\tt OpenMP} configuration that leads to a greenup (i.e., greenup = $\frac{Energy_{old}}{Energy_{new}}$~\cite{choi2013roofline}) of $3.89\times$, but a speedup of $0.95\times$ (i.e., a \emph{slowdown}) over the typical {\tt OpenMP} configuration at TDP ($85$W).
This contradicts the commonly held belief of \emph{race-to-halt}~\cite{AWAN2014796} (i.e., the idea that the lower energy consumption occurs at the highest speedup), and shows that optimizing for time and optimizing for energy may not yield the same {\tt OpenMP} configuration.
In addition, for applications where a slowdown is unacceptable, we can simultaneously optimize for time and energy by targeting the energy-delay product (EDP) metric~\cite{laros2013energy}.
Through an exhaustive search through the {\tt OpenMP} configurations space, we observe that minimizing EDP yields a speedup of $1.64\times$ and a greenup of $2.7\times$, at a yet another {\tt OpenMP} configuration and power constraint.

In summary, optimizing for performance, power, and energy consumption all require different strategies for identifying the optimal {\tt OpenMP} configuration, and optimizing for one metric (e.g., performance) does not necessarily optimize for another (e.g., energy).
To this end, we propose a graph neural network (GNN)-based technique that can be used to 
\begin{enumerate*}[label=(\roman*)]
    \item identify {\tt OpenMP} configurations at prescribed power constraints that maximizes performance and
    \item optimize for the \textit{energy-delay product} to identify configurations for both energy-efficient and performant execution
\end{enumerate*}.

In this study, {\tt OpenMP} code regions are first transformed to a flow-aware graphical representation. 
These code graphs are then modeled by a GNN, and used for predicting the best configurations for the appropriate target. 
In contrast to prior studies, we use only these code graphs (i.e., static features) as inputs to our model, which does not require \emph{expensive code execution}.
The benefit of using a deep learning (DL)-based approach is that it automatically helps reduce the search space exploration by aggressively pruning non-beneficial points in the search space.

The works in \cite{bari2016arcs, bari2019performance} studied the impacts of power constraints and {\tt OpenMP} configurations on time and energy and are, to the best of our knowledge, most similar to the problem considered in this paper.
To demonstrate the effectiveness of our static approach, we compare our results against a Bayesian Optimization based tuner {\tt BLISS} \cite{roy2021bliss}, and a search-based tuner {\tt OpenTuner} \cite{ansel2014opentuner}.
Through this study, we propose two separate approaches for tuning performance and energy/power. 
The first approach aims to identify the tuning configuration that can produce the fastest execution at a predefined power constraint.
The second approach looks at both time and energy as target metrics and aims to optimize for both at the same time by identifying configurations that lead to the lowest \textit{energy-delay product}. 
The key contributions of our work are as follows:
\begin{itemize}[leftmargin=*]
    \item We build an RGCN network to model flow-aware {\tt OpenMP} code region graphs that captures both semantic and structural features of code regions, and is portable across different architectures.
    \item We build an auto-tuning framework that identifies {\tt OpenMP} configurations yielding near optimal execution times at different power constraints.
    We achieve a geometric mean speedup of $1.33\times$ and $1.15\times$ over default {\tt OpenMP} configurations at four power constraints across $30$ applications on Skylake and Haswell systems.
    \item Our DL-based framework also optimizes for both time and energy simultaneously by minimizing the EDP.
    We achieve geometric mean speedup of $1.27\times$ and $1.12\times$, and greenup of $1.40\times$ and $1.22\times$ respectively on Skylake and Haswell, over default {\tt OpenMP} configurations running at TDP (i.e., no power constraint).
    \item We compare our framework against the state-of-the-art {\tt BLISS}~\cite{roy2021bliss} tuner and {\tt OpenTuner} \cite{ansel2014opentuner} and demonstrate better performance without the need for executing code.
\end{itemize}
\section{Background}
This section outlines concepts relevant to this work.
\subsection{Static Code Representations for DL}
\label{sec:bg_code_rep}
DL is being increasingly used for code analysis and optimization tasks \cite{allamanis2018survey}. However, the use of DL necessitates the use of a strong code representation capable of capturing the inherent features in source code.
A lot of prior studies have represented programs as a sequence of lexical tokens \cite{cummins2021programl}. 
But, these fail to capture the structured nature of programs. 
To overcome this, representations capturing syntactic as well as semantic features have been proposed \cite{allamanis2018survey, brauckmann2020compiler} . 

These methods often do not take into account control, data, or call flows in the program. 
PROGRAML \cite{cummins2021programl} is a tool that represents the semantic and structural features of code in a flow-aware multi-graph.
These DL-friendly multi-graphs have a vertex for each instruction and control-flow edges between them. Data flow is represented by separate vertices for variables and constants and associated data-flow edges to instructions. Call flow is represented by edges between callee functions and caller instruction vertices. We use this tool to transform code region IRs to their corresponding graphs.

\subsection{Power Constraining and Energy Profiling}


Starting with the SandyBridge $\mu$architecture, Intel introduced the RAPL software tool that enables power/energy monitoring and power capping through a simple interface.
The power to several subsytems of the processor, such as memory, DRAM, CPU, etc can be controlled via RAPL. 
We use the \textit{Variorum} \cite{brinkvariorum} tool, which in turn uses RAPL and device MSRs, to control the power constraint on the CPU.
We also use PAPI (with the RAPL component enabled) \cite{mucci1999papi} to measure performance counters and energy profiling data.

\subsection{Graph Neural Networks}
\label{background:gnn}
Recent advances in deep learning have now enabled the application of DL on data generated from non-Euclidean space \cite{wu2020comprehensive}.
The relations and dependencies between objects in such data can more readily be represented as a graph. GNNs were proposed as a means of modeling such data.
Graph Convolutional Networks (GCNs) are a form of GNNs aimed at generalizing the common \textit{sliding window} convolution operation on grid data in regular Convolutional Neural Networks to graphs \cite{wu2020comprehensive}. A GCN network updates its node representation by aggregating the features from the node's neighbors along with the node. Similar to CNNs, GCNs stack multiple convolutional layers to extract high-level node representation. 
We use Relational Graph Convolutional Network (RGCN), a variation of GCN, to model our program graphs.
RGCNs were proposed to enable networks to better model large-scale relational data \cite{schlichtkrull2018modeling}. RGCNs differ from GCNs in that they work with relation specific transformations annotated by the type and direction of edges. RGCNs accumulate transformed feature vectors through a normalized sum.
\section{The PnP Auto-Tuner: A GNN based Power and Performance Tuner}
\begin{figure*}
    \centering
    \includegraphics[ width=0.95\textwidth]{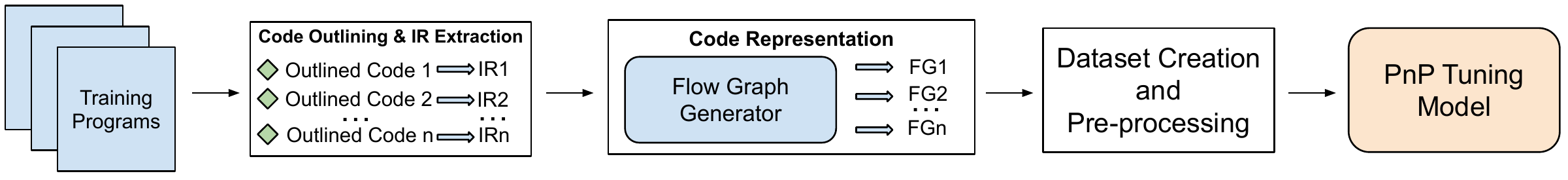}
    \caption{PnP Tuner Pipeline: An overview of tasks in our GNN based power and performance tuner}
    \label{fig:pnp_pipeline}
\end{figure*}
In this section, we outline our two-pronged approach to tuning performance and power.
We consider two scenarios with real-world implications: i) Because of cost and energy considerations, clusters and data-centers must usually work under strict power budgets.
However, constraining power directly impacts performance by limiting the power delivered to hardware components.
Therefore, assuming no code changes or compiler optimizations, tuning available runtime parameters becomes essential for improving application performance.
ii) It is of utmost importance in most HPC systems to reduce energy consumption.
This has a direct monetary and environmental impact.
However, as shown in Section \ref{sec:introduction}, simply optimizing for energy, can potentially lead to slower executions.
Therefore, we must optimize for a metric that considers both energy and performance.
To this end, we target the multi-obejctive metric \textit{energy-delay product (EDP)}.
We use GNNs to build a model that will be used for the aforementioned tasks.
The inputs to the GNNs are code flow graphs of {\tt OpenMP} regions.
Using such graphs allows us to model the semantics and structure of source code.
These convey relevant information to the model about the code region being tuned.
We refer to these input code graphs as static features, as these are obtained statically without any code executions.
An overview of this pipeline is shown in Figure \ref{fig:pnp_pipeline}, and outlined in the following paragraphs.


\subsection{Representing the Code}
\label{sec:code_representation}
In this study, we aim to optimize {\tt OpenMP} code regions. 
These code regions are usually the primary computational bottlenecks in such applications. 
Instead of focusing on individual loops inside these parallel regions, we aim to optimize the parallel region as a whole for larger performance improvements. 
Tuning sub-regions within an {\tt OpenMP} code region adds additional overhead.
Switching between configurations can improve the performance of each sub-region (loops for example), but can degrade the performance of each {\tt OpenMP} region and the application as a whole.
The benchmark applications are initially compiled to their intermediate representations (IR). 
Compiling {\tt OpenMP} code to its corresponding IR automatically encloses the parallel region in an outlined function. 
We use the \texttt{llvm-extract} tool to extract the outlined parallel region. 
As shown in Figure \ref{fig:pnp_pipeline}, to represent the code regions in a form usable by DL models, we use PROGRAML \cite{cummins2021programl} to obtain the corresponding graph embeddings. These code graphs encapsulate the semantic and structural characteristics of code, as well as the data flow, control flow, and call flow in source code, as described in Section \ref{sec:bg_code_rep}.

\subsection{Configuring the Search Space}
One of the primary motivations behind using a DL technique for this work was to develop a method that can work with large search spaces easily. 
Unlike most existing auto-tuners, which have to extensively execute programs to identify the best configurations, our DL-based framework will not need to execute programs.
For the proposed DL approach to scale well to unseen code and inputs, it is necessary to feed the model with code graphs with enough variability.
Along with variability in considered parallel code regions, it is essential to model the effect of various tuning parameters on these code regions.
Different configurations impact code execution by affecting the load balancing and cache behavior, which in turn impacts performance. 

As our goal is to target performance optimization and energy efficiency, we must simultaneously consider the impact of power constraints and {\tt OpenMP} parameters on code executions. 
To this end, we have defined a search space (shown in Table \ref{tab:search-space}) with $504$ valid configurations. 
In addition, the default {\tt OpenMP} configurations for each of the four power limits have also been considered as valid configurations leading to a total of $508$ configurations.
The search space used in this study has been selected based on ideas presented by Bari et al. in \cite{bari2019performance}.

\begin{table}[htbp]
\centering
\caption{Search space for performance and power tuning on Skylake and Haswell nodes.}
\begin{tabular}{p{0.3\linewidth}p{0.55\linewidth}}
\hline
\textbf{Search Space} & \textbf{Parameter Values} \\
\hline
Power Limits & 75W, 100W, 120W, 150W (Skylake) \newline
40W, 60W, 70W, 85W (Haswell)
\\
Number of threads & 1, 4, 8, 16, 32, 64 (Skylake)
\newline
1, 2, 4, 8, 16, 32 (Haswell)\\
Scheduling Policy & STATIC, DYNAMIC, GUIDED\\
Chunk Sizes & 1, 8, 32, 64, 128, 256, 512\\
\hline 
\end{tabular}
\label{tab:search-space}
\end{table}
\subsection{Power Constraining and Dataset Creation }
In this work, we used the \textit{Variorum} \cite{brinkvariorum} tool for constraining power levels on each of the experimental systems. 
We used Variorum APIs to interface with RAPL and device MSRs to constrain power to the values described in Table \ref{tab:search-space}.

To validate our hypothesis, we chose to work with multiple {\tt OpenMP} applications with varied complexity.
These {\tt OpenMP} regions consists of parallel regions with simple do-all loops to regions with multiple loops with varying levels of nesting and diverse programmatic constructs.
We have worked with $25$ applications from the PolyBench suite \cite{pouchet2012polybench}, and mini and proxy applications XSBench \cite{tramm2014xsbench}, RSBench \cite{tramm2014performance}, miniFE \cite{hammondminife}, miniAMR \cite{sasidharan2016miniamr}, Quicksilver \cite{quicksilverllnl}, and LULESH \cite{karlin2013lulesh} with combined total of 68 {\tt OpenMP} regions.

At each power level, parallel {\tt OpenMP} regions in all considered applications were executed for each runtime configuration in Table \ref{tab:search-space} and default {\tt OpenMP} configurations (all threads, static scheduling, and compiler defined chunk sizes) on each system.
The execution times obtained as such are then analyzed to identify the best configuration for each code region.
The best configurations are used as labels during training.

\subsection{Performance and Power Modeling}
This section outlines our GNN-based approach towards performance and power optimizations. 
We propose two tuning scenarios with different objectives:
\begin{itemize}
    \item In the first scenario, we aim to identify the {\tt OpenMP} configuration that lead to the fastest executions at a given power constraint.
    \item In the second scenario, we aim to identify both the {\tt OpenMP} configuration and the power level that minimizes the EDP.
    By minimizing the EDP, we hope to improve the execution time \textit{and} energy efficiency in comparison to default {\tt OpenMP} configuration at TDP.
\end{itemize}

\subsubsection{Code Graph Modeling using GNNs} 
\label{sec:code_modeling}
For both scenarios, the code modeling technique is similar.
Modeling code graphs allows us to model code semantics and structure.
Analyzing code structure allows us to better capture the interdependence between code blocks.
Simply looking at code as a sequence of text does not afford this information.
The code graphs generated in Section \ref{sec:code_representation} are initially passed through a GNN network for modeling the code graphs. Specifically, Relational Graph Convolutional Networks (RGCNs) are used as these allow modeling relation specific features. 
Each code graph consists of three types of edges denoting the type of flow (Section \ref{sec:code_representation}).
The type of edges are used as edge features during modeling.
For each node in a graph, the node features are the type of node, and the associated IR code block.
Before modeling, the code region IRs are used to generate an embedding.
This embedding maps IR text to tensors.
These tensors are then passed to the model as node features along with the type of the node.
Based on these features, the GNN layers model these by passing ``messages'' between neighboring nodes, aggregation, and subsequent
weight updations \cite{zhou2020graph}.
The output tensors from the GNN layers then fed into fully connected neural network layers with the aim of identifying the best configurations.

\subsubsection{Power Constraint Specific Auto-tuning}
\label{sec:pow_const_tuning}
As noted in Section \ref{sec:introduction}, one way of meeting power consumption goals is to enforce a specific power constraint. 
Such power constraints can help limit the power drawn by a node or its subsystems. 
However, simply using default {\tt OpenMP} runtime configurations at different power constraints for code execution may lead to performance degradation, as well as increased energy usage from static power.
Therefore, we aim to identify those configurations that lead to speedups at predefined power constraints. 
We propose a DL based technique for power-constrained auto-tuning. 
As outlined in Section \ref{sec:code_modeling}, we use the flow-aware code graphs obtained from the parallel code region IRs as inputs to the RGCN layers of our network.
As shown in Figure \ref{fig:pnp_pipeline}, the RGCN layers model each such graph and feeds the output into a fully connected (dense) network. 
The dense layers acts as a classifier and are trained as such with the target of predicting the best configuration for a given {\tt OpenMP} code region.


\subsubsection{Optimizing Energy and Time}
For nodes and systems without any predefined power constraint, time and energy optimization are still of primary importance. 
However, simply optimizing for performance \emph{or} energy neglects the other criteria.
Thus, in this section, we propose using power constraints as a tuning parameter along with the available {\tt OpenMP} runtime configurations for joint optimization of performance and power.
Simply using execution time or energy savings for identifying such configurations is not enough. 
Thus, we use the \textit{energy-delay product (EDP)} metric \cite{laros2013energy} as a more accurate measure of the impact of different configurations on code performance. 
In this work, we assign equal importance to time and energy and use the metric $E *  T$, where $E$ represents the energy consumption, and $T$ represents the execution time for a parallel code region.

\begin{figure*}
\begin{tabular}{c}
 \includegraphics[width=\textwidth]{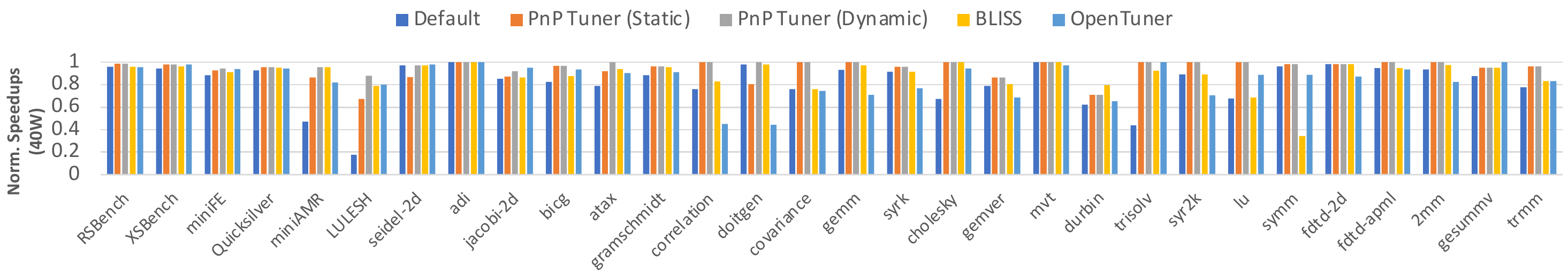}\\
 \includegraphics[width=\textwidth]{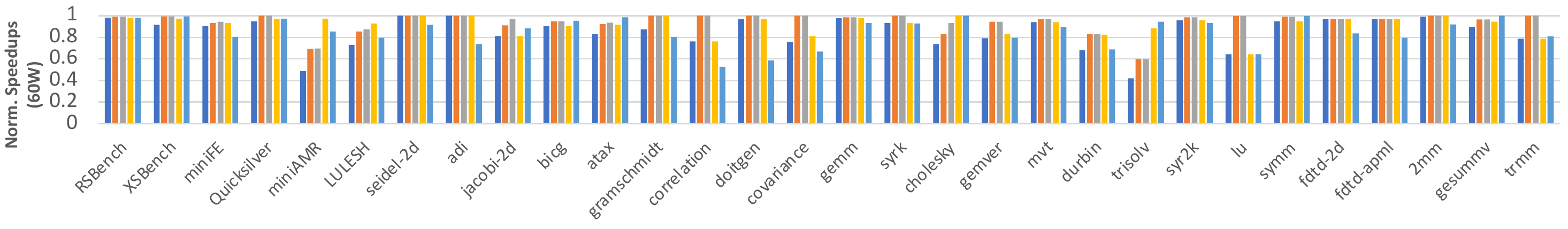}\\
 \includegraphics[width=\textwidth]{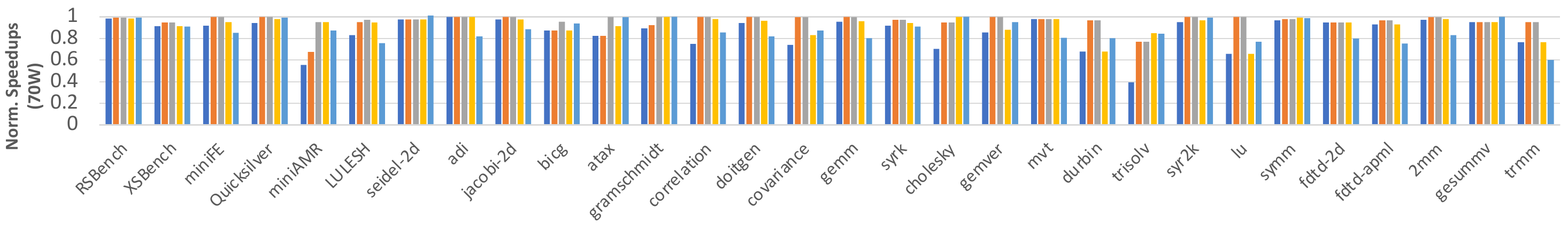}\\
 \includegraphics[width=\textwidth]{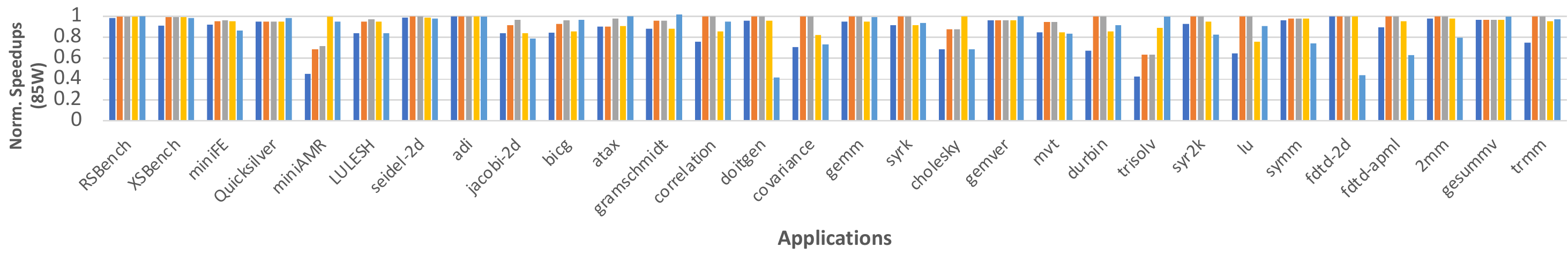} \\

\end{tabular}
\caption{Power Constrained Tuning (Haswell): Each chart shows results for a specific power constraint.
Each bar-group shows geometric mean speedup for all {\tt OpenMP} regions in an application over default {\tt OpenMP} settings wrt the corresponding tuning approach.
Speedups are normalized by oracle(brute-force) speedups. 
Normalized oracle speedups are always $1.0\times$.
The PnP tuner outperforms {\tt BLISS} in $82.5\%$ and {\tt OpenTuner} in $78\%$ cases across all power constraints(see Section~\ref{sec:pwr_const_tuning_exp} for details).}
\label{fig:pwr_const_tun_speedup_haswell}
\end{figure*}

We again use the modeled code graphs from Section \ref{sec:code_modeling} as the static feature inputs to our model for this experiment and train our model with a target of optimizing the EDP.
As in the previous subsection, a fully connected neural network serves as a classifier to identify the best configurations for tuning EDP. 
Using a DL-based approach for identifying the best one out of $508$ possible configurations is especially beneficial, as such models are efficient at automatically pruning the under performing configurations. 
This is in stark contrast to brute-force approaches, where the tuning cost would explode with increasing search space complexity.

\section{Experiments}
To identify near optimum values of tuning parameters for both our experimental scenarios, we first explore every permutation of inputs and configurations considered in this study. 
We use this exhaustive exploration as an oracle to compare the results from our work. 
We also compare our work against {\tt BLISS} \cite{roy2021bliss} and {\tt OpenTuner} \cite{ansel2014opentuner}.
All results presented in the following paragraphs represent speedups/greenups of each code region.
For applications with multiple {\tt OpenMP} regions, the geometric mean of speedups/greenups of all regions in an application are reported.
We have also verified that there are sequences of serial code in between successive {\tt OpenMP} regions.
This allows us to look at each region as a self-contained unit, and makes them good candidates for tuning.
We assume that the performance of these intervening serial sequences will not change and improving the performance of each {\tt OpenMP} region would translate to improvement in application performance.
\subsection{Experimental Setup}
For our experiments, we use two systems; one with Intel(R) Xeon(R) Gold $6142$ CPU with $32$ cores, two hyper-threads per core, and two sockets (Skylake) with a minimum and TDP package power of $75W$ and $150W$, and an Intel(R) Xeon(R) E5-2630 v3 CPU (Haswell), with 16 cores, two hyper-threads per core, and two sockets, and minimum and TDP package power of $40W$ and $85W$.
We use Clang tools for code compilation and transformation to IR, and {\tt PyTorch} DL libraries for building our GNN models.

\subsection{Power Constrained Auto-tuning}
\label{sec:pwr_const_tuning_exp}

\begin{figure*}
\begin{tabular}{c}
 \includegraphics[width=\textwidth]{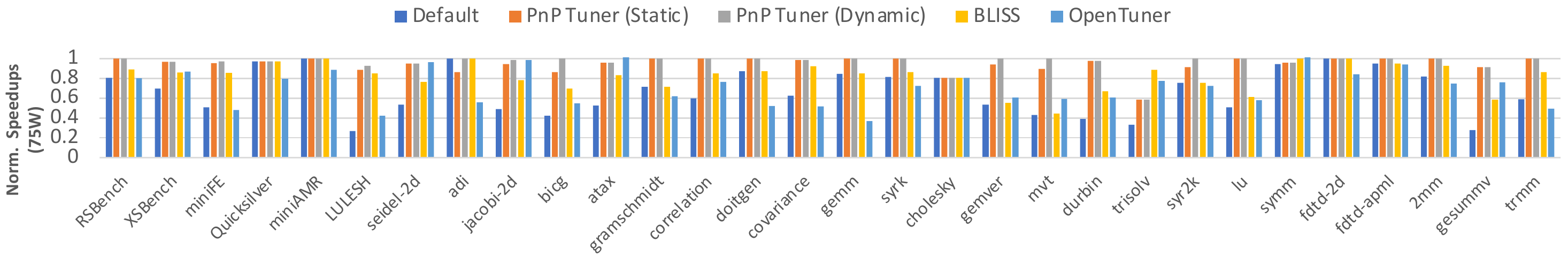}\\ 
 \includegraphics[width=\textwidth]{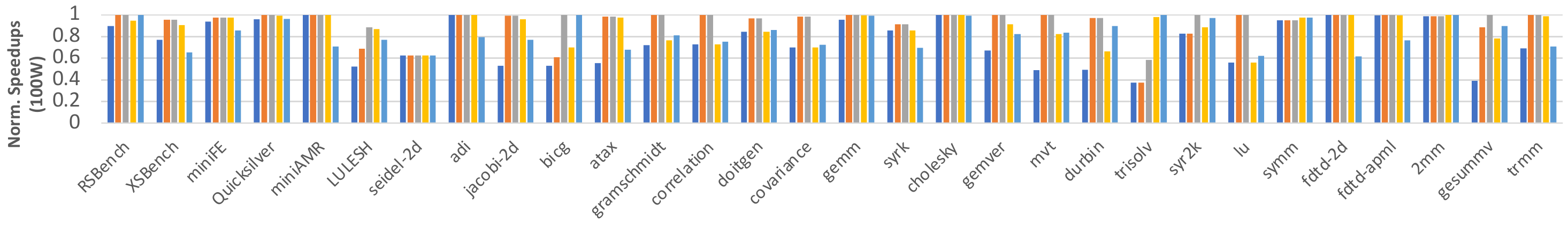}\\
 \includegraphics[width=\textwidth]{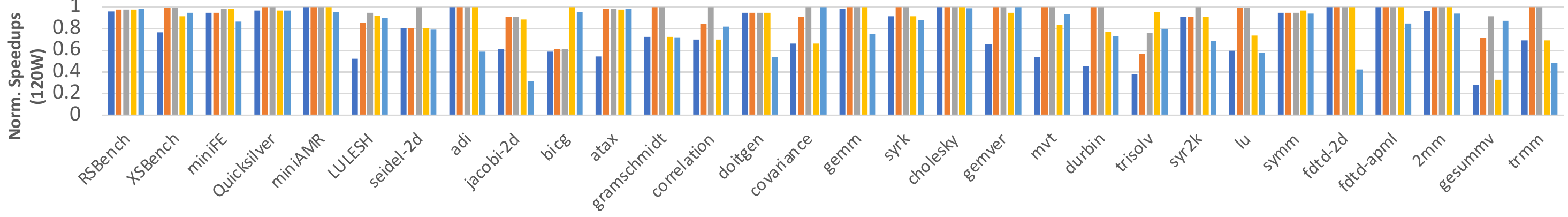}\\ 
 \includegraphics[width=\textwidth]{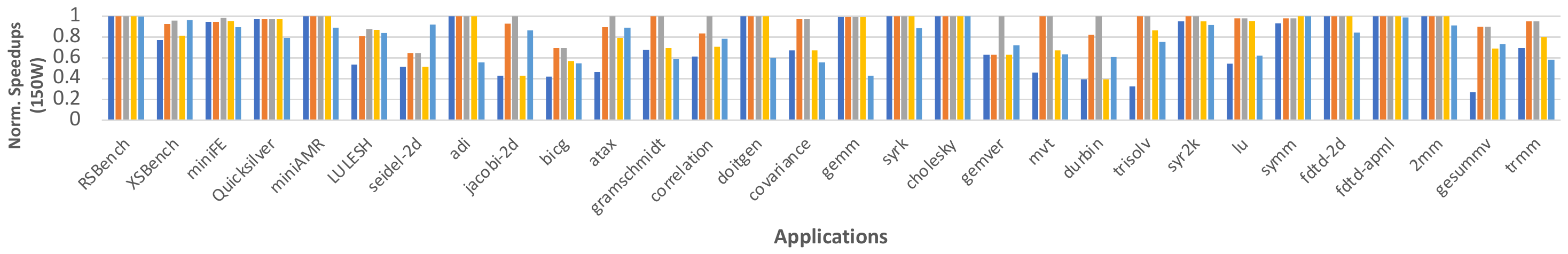} \\

\end{tabular}
\caption{Power Constrained Tuning (Skylake): Each chart shows results for a specific power constraint.
Each bar-group shows geometric mean speedup for all {\tt OpenMP} regions in an application over default {\tt OpenMP} settings wrt the corresponding tuning approach.
Speedups are normalized by oracle speedups. 
Normalized oracle speedups are always $1.0\times$.
The PnP tuner outperforms {\tt BLISS} in $85\%$ and {\tt OpenTuner} in $83\%$ cases across all power constraints(see Section~\ref{sec:pwr_const_tuning_exp} for more details).}
\label{fig:pwr_const_tun_speedup_skylake}
\end{figure*}

In this section, we evaluate the performance of our tuner in determining the optimal configuration for minimizing execution time given a specific power constraint (described in Section~\ref{sec:pow_const_tuning}).
To validate the effectiveness of our approach, we use \textit{leave-one-out cross-validation (LOOCV)}. 
For each fold, code regions from one benchmark application is selected and assigned to the validation set and the code regions from all other applications are assigned to the training set. 
We repeat this process for all applications in our approach. 
Such a process is essential to evaluate the performance of our model on previously unobserved code regions.

The results for the Haswell system are shown in Figure \ref{fig:pwr_const_tun_speedup_haswell}. 
For each application, we calculate the geometric mean speedups for all {\tt OpenMP} regions in each application achieved by each tuner across four power constraints (i.e., $40$W, $60$W, $70$W, $85$W).

While training the model on the data from the Skylake system, we borrow ideas from transfer/inductive learning and perform an optimization step to speed up the training process.
Because the code graphs are statically generated, the code graphs obtained on different systems using the same compiler are identical.
For this reason, we save the weights and model states of the GNN model obtained while training our model on the Haswell system.
While training the model on the Skylake data, we load the saved weights and model and only re-train the dense layers.
This leads to $4.18\times$ faster training (or reduces training time by $76\%$).

Results for each power constraint ($75$W, $100$W, $120$W, $150$W) is shown in Figure \ref{fig:pwr_const_tun_speedup_skylake} for the Skylake system.
Each speedup is normalized by the speedup achieved by the oracle (i.e., exhaustive exploration).
In $74\%$ cases (across both systems and power constraints), our PnP tuner identifies configurations that lead to $>=0.95\times$ of the oracle speedups (assuming oracle as $1.0\times$). 
These results are obtained without executing the code.
In contrast, {\tt BLISS} and {\tt OpenTuner} needs to execute code multiple times and achieves $>=0.95\times$ of the oracle speedups in $51\%$ and $34\%$ cases for {\tt BLISS} and {\tt OpenTuner} respectively.
The PnP tuner produces better results than {\tt BLISS} and {\tt OpenTuner} in $83\%$ and $78\%$ cases.
Overall, the configurations predicted by our model lead to geometric mean speedups of $1.19\times$, $1.12\times$, $1.13\times$, and $1.14\times$ for power limits $40W$, $60W$, $70W$, and $85W$ on the Haswell system.
In contrast, {\tt BLISS} leads to speedups of $1.11\times$, $1.09\times$, $1.09\times$, and $1.11\times$ across these power constraints respectively.
{\tt OpenTuner} produces corresponding speedups of $1.06\times$, $1.0\times$, $1.04\times$, and $1.02\times$.
On Skylake, our approach achieves geometric mean speedups of $1.5\times$, $1.25\times$, $1.26\times$, and $1.34\times$ across power constraints $75W$, $100W$, $120W$, and $150W$ respectively, compared to speedups of $1.29\times$, $1.2\times$, $1.18\times$, and $1.17\times$ produced by {\tt BLISS}, and speedups of $1.27\times$, $1.13\times$, $1.07\times$, and $1.1 \times$ produced by {\tt OpenTuner}.

\noindent\textbf{\textit{Can performance counters further improve results?}}
\label{sec:pwr_const_tuning_perf_impact}
Although our approach leads to $>=0.95\times$ of the oracle speedups in most cases, in approximately $8\%$ of cases, our approach produces results which are $<0.8\times$ of the oracle speedups.
Previous works such as \cite{alcaraz2022predicting, sanchez2020modeling} have used performance counters for tuning tasks.
We borrow from these ideas to see if the results from our approach can be improved by using these as features (dynamic features).
For this experiment, we update our model definition.
We make no changes to the GNN layers.
We repurpose the fully connected layers to accept as inputs five performance counters along with the ouputs from the GNN layers.
We use PAPI \cite{mucci1999papi} to collect counters related to L1, L2, L3 cache misses, number of instructions, and the number of mispredicted branches for each {\tt OpenMP} region.
These were selected as these have direct impact on code execution and performance.

We perform the same experiments as outlined in the previous paragraphs.
However, we only validate on those applications whose speedups are $<0.95\times$ on the oracle speedups.
We see that by including performance counters, this approach identifies configurations that lead to $>=0.95\times$ in $87.5\%$ cases (up from $74\%$).
We show these results and comparisons in Figures \ref{fig:pwr_const_tun_speedup_haswell} and \ref{fig:pwr_const_tun_speedup_skylake}.
Therefore, a case can definitely be made for including performance counters for DL-based performance tuning.
However, this comes at the additional cost of profiling.
Profiling is necessary for generating the dataset to train the model.
However, during inference, this approach (using both static and dynamic features) only needs to execute applications twice (to collect counters which serve as inputs to the model), which is less than other execution based tuners.
To conclude, although this produces better results, it adds a profiling overhead.
But during inference, in spite of this overhead it only needs two executions.

\noindent\textbf{\textit{Can we extend this approach to unknown power constraints?}}
There might be scenarios where adding/removing new nodes to/from clusters, or other factors, might necessitate changing power constraints on nodes.
Thus, our approach should also be generalizable to power constraints that our model has not been trained on, since data center policy changes may result in different power constraints being applied.
To evaluate this scenario, we conduct four tests - two tests for each system - one test each for the lowest and highest power constraints considered in this paper.
For each test, we first \emph{exclude} all measurements for the target power constraint (e.g., for the $150$W test on Skylake, for training, we use measurements from $75$W,  $100$W, and $120$W only).
We then train and validate our model using \textit{leave-one-out cross-validation} as before.
This allows us to generalize for both unseen applications \emph{and} unseen power constraints.
However, unlike the initial experiments which uses a \textit{static-only} approach, we use performance counters as part of the feature set in this experiment.
This is to account for the variation in runtime behavior of parallel regions under varying power constraints.
Static features cannot encapsulate such divergence in behavior. 
The input features and model is similar to the one described in Section \ref{sec:pwr_const_tuning_perf_impact}.
In addition to these features, we also input as feature the normalized power constraints for each feature set.
This helps to associate runtime behavior (performance counters) with power limits.
\begin{figure}
    \centering
    \includegraphics[width=0.49\textwidth]{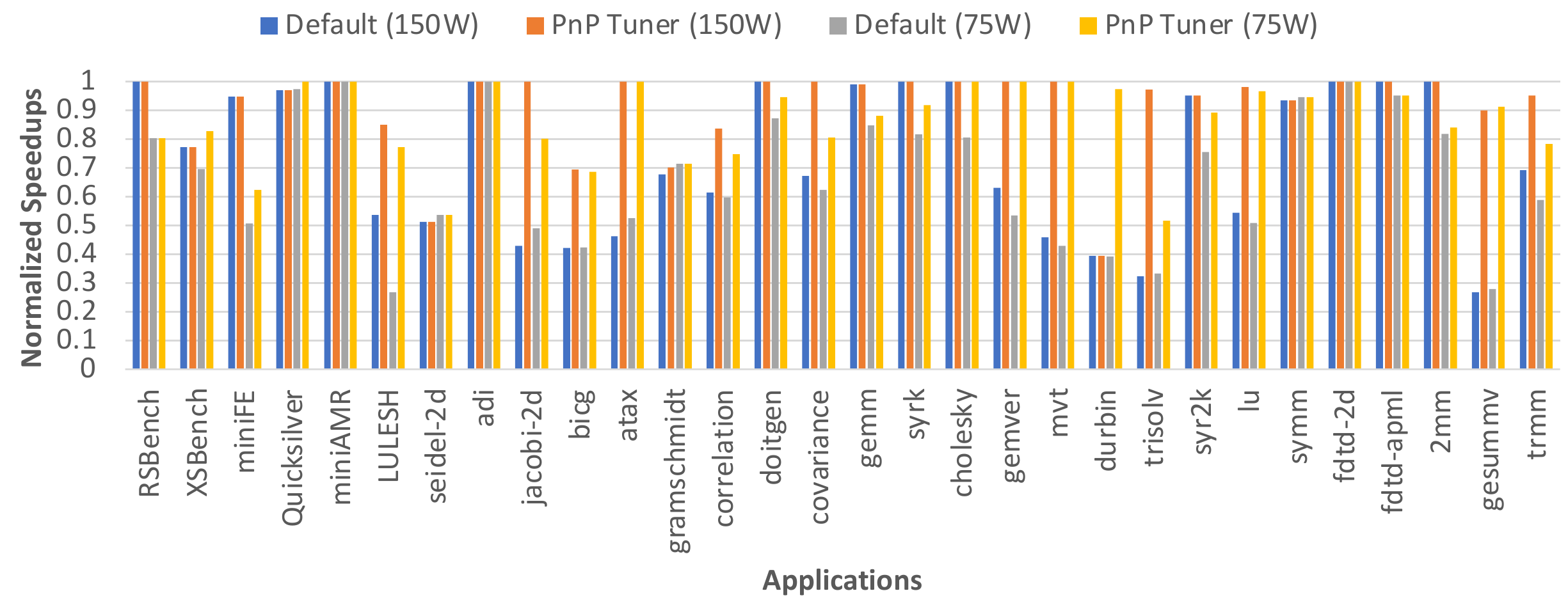}
    \caption{Power Constrained Tuning on unseen power constraints (Skylake): Geometric mean speedup over default {\tt OpenMP} settings.
    Results normalized by the oracle speedup.}
    \label{fig:pwr_const_tun_unseen_clemson}
\end{figure}

\begin{figure}
    \centering
    \includegraphics[width=0.49\textwidth]{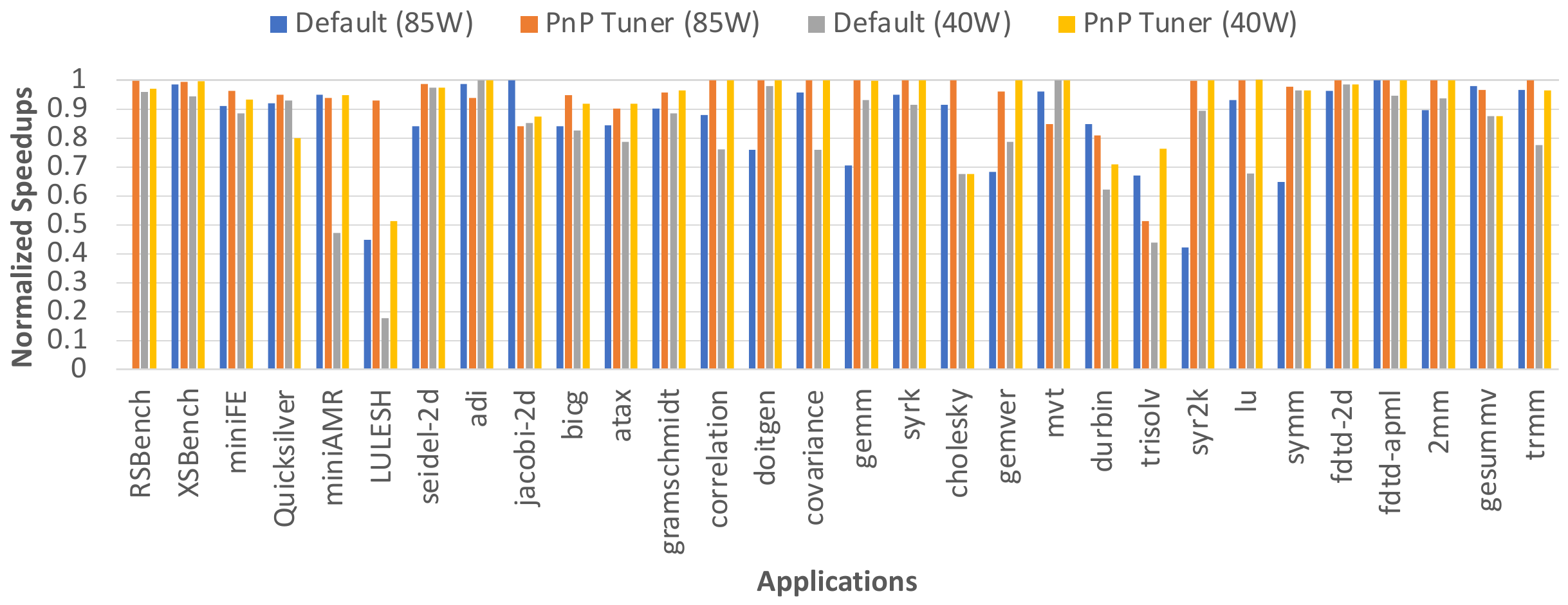}
    \caption{Power Constrained Tuning on unseen power constraints (Haswell): Geometric mean speedup over default {\tt OpenMP} settings.
    Results normalized by the oracle speedup.}
    \label{fig:pwr_const_tun_unseen_haswell}
\end{figure}

Figures~\ref{fig:pwr_const_tun_unseen_clemson} and \ref{fig:pwr_const_tun_unseen_haswell} shows that our model performs well in such scenarios for both the Skylake and Haswell systems, predicting configurations that are within $5$\% (i.e., $\ge$ $0.95$ normalized speedup) of the best possible speedup in $64\%$ cases and within $20\%$ of the best possible speedups in $85\%$ cases across both systems and four power constraints.
On the Skylake systems, these tuning efforts lead to geometric mean speedups of $1.29\times$ and $1.36\times$ versus oracle speedups of $1.44\times$ and $1.59\times$ for power constraints of $150$W and $75$W respectively.
On the Haswell system, these experiments produce speedups of $1.13\times$ and $1.17\times$ compared to oracle speedups of $1.16\times$ and $1.27\times$ for power constraints of $85$W and $40$W respectively.

\begin{table}[htbp]
\centering
\caption{Deep Learning Model Hyperparameters.}
\begin{tabular}{p{0.25\linewidth}p{0.6\linewidth}}
\hline
\textbf{Hyperparameter} & \textbf{Hyperparameter Values} \\
\hline
Layers & RGCN (4), FCNN (3)\\
Activ. func. & Leaky ReLU, ReLU\\
Optimizer & AdamW ({\small amsgrad}) ({\small Sec \ref{sec:pwr_const_tuning_exp}}), Adam ({\small Sec \ref{sec:power_perf_tuning}})\\
Learning Rate & 0.001\\
Batch Size & 16\\
Loss function & Cross Entropy Loss\\
\hline 
\end{tabular}
\label{tab:ml-params}
\end{table}

The hyperparameters of the models used in these experiments are shown in Table \ref{tab:ml-params}. Other parameter values may have minor differences between experiments.
\subsection{Power and Performance Tuning}
\label{sec:power_perf_tuning}
With increasing financial and environmental impacts of high energy usage, energy efficiency is now as important as performance in the current HPC landscape.
However, simply optimizing for energy consumption, as shown in Section \ref{sec:introduction}, may lead to lower performance. 

\begin{figure*}
\begin{tabular}{cc}
  \includegraphics[width=0.5\textwidth]{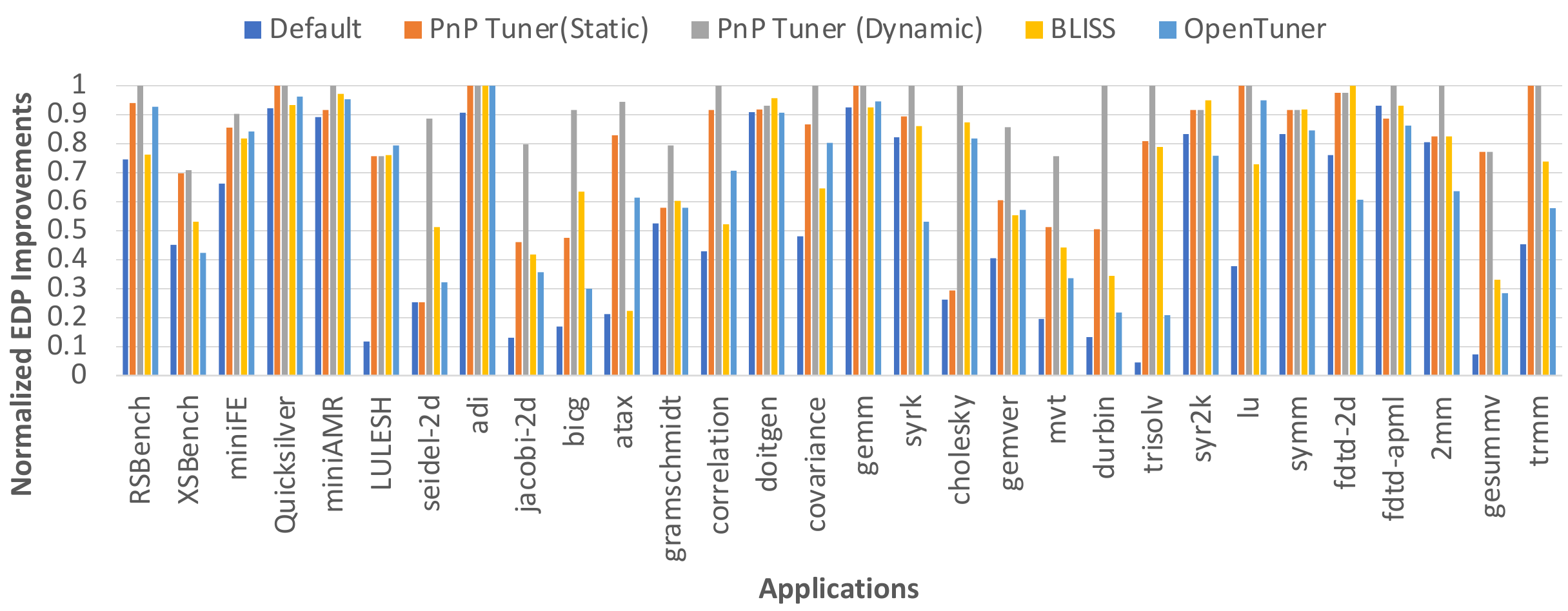} &   \includegraphics[width=0.5\textwidth]{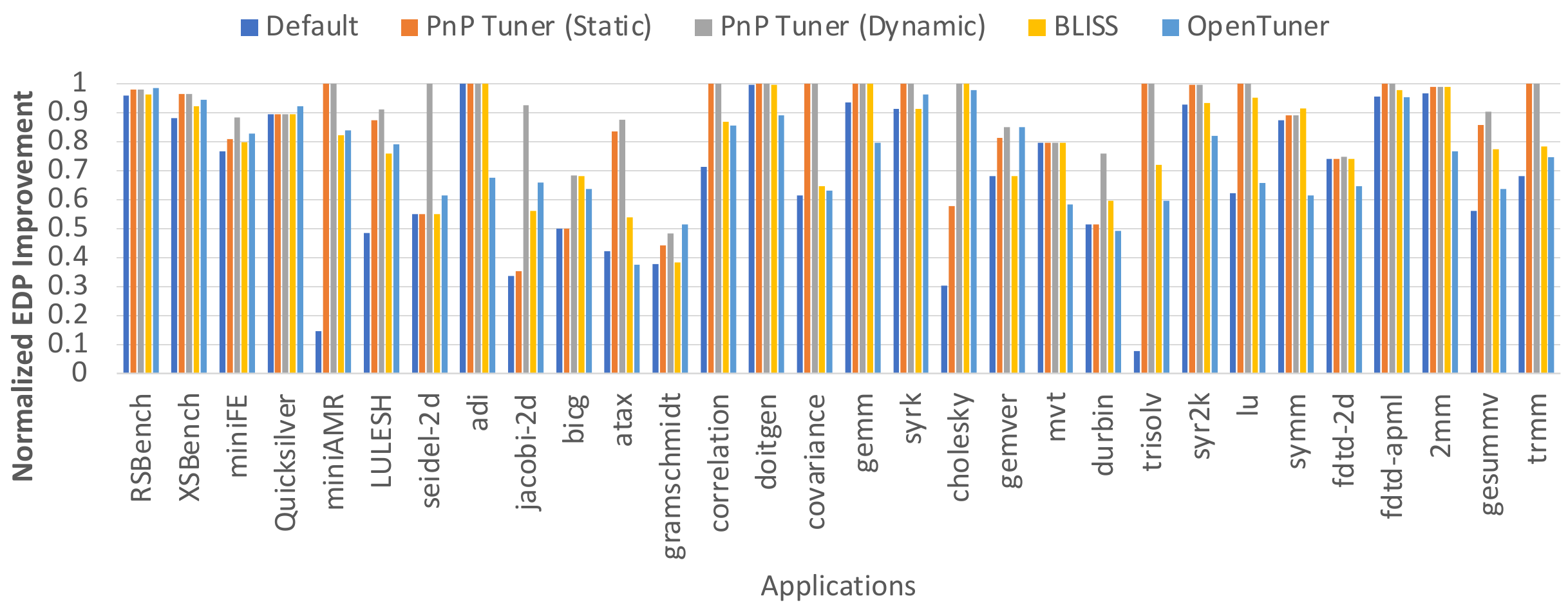} \\
(a) Skylake & (b) Haswell

\end{tabular}
\caption{Improvement in EDP over default {\tt OpenMP} configurations for each application for the Haswell system. EDP improvements normalized in terms of best achievable EDP improvement.}
\label{fig:edp_improv}
\end{figure*}

Thus, in this section, we outline the second scenario mentioned at the beginning of this section.
To this end, we build a GNN-based tuner that uses only static features, with the aim of identifying a combination of power constraints and {\tt OpenMP} runtime configurations that can lead to performance improvement while reducing energy consumption. 
As in the previous experiments, we model our flow-aware code graphs using an RGCN network. 
The outputs from the GNN layers are fed into the dense layers.
These layers are trained with the target of finding configurations that produce the best energy-delay product (EDP). 
Again, we use \textit{leave-one-out cross validation} to validate our model, and the process of assigning benchmark applications to the training and validation set is similar to that described in Section \ref{sec:pwr_const_tuning_exp}. 

The configurations predicted in these experiments lead to within $5\%$ of the oracle EDP improvements in $45\%$ cases, and within $20\%$ of the oracle improvements in $69\%$ cases across the two systems.
In comparison, {\tt BLISS} reaches these numbers in $35\%$ and $45\%$ cases (Figure \ref{fig:edp_improv}).
{\tt OpenTuner} reaches these numbers in $22\%$ and $40\%$ cases.
Overall, the configurations predicted by our \textit{static-only} approach leads to geometric mean improvements of $1.37\times$ and $1.85\times$ on the Haswell and Skylake systems compared to $1.31\times$ and $1.69\times$ respectively achieved by {\tt BLISS} and $1.21\times$ and $1.49\times$ achieved by {\tt OpenTuner}.
\begin{figure*}
\begin{tabular}{c}
  \includegraphics[width=\textwidth]{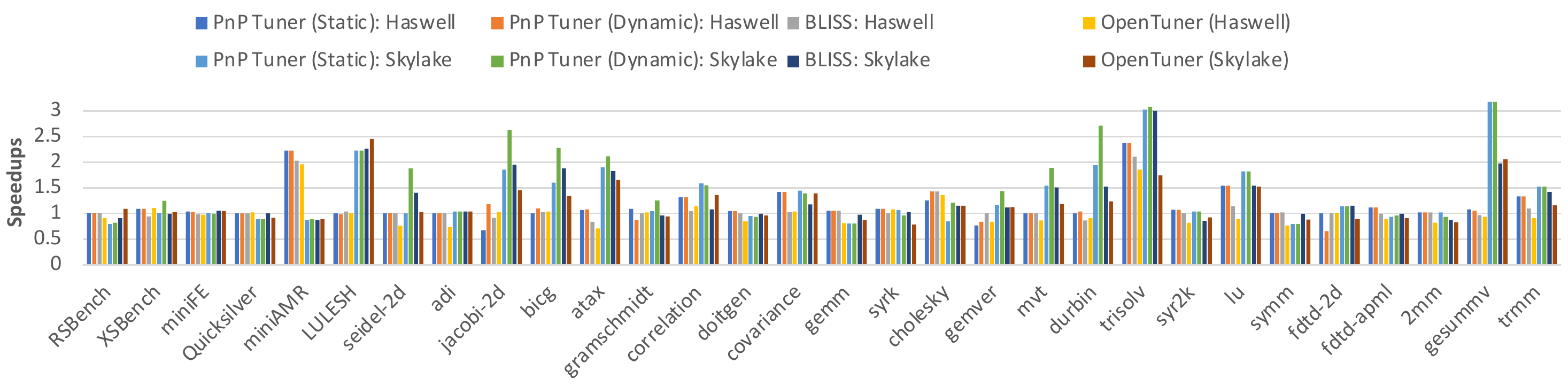}\\ 
  \includegraphics[width=\textwidth]{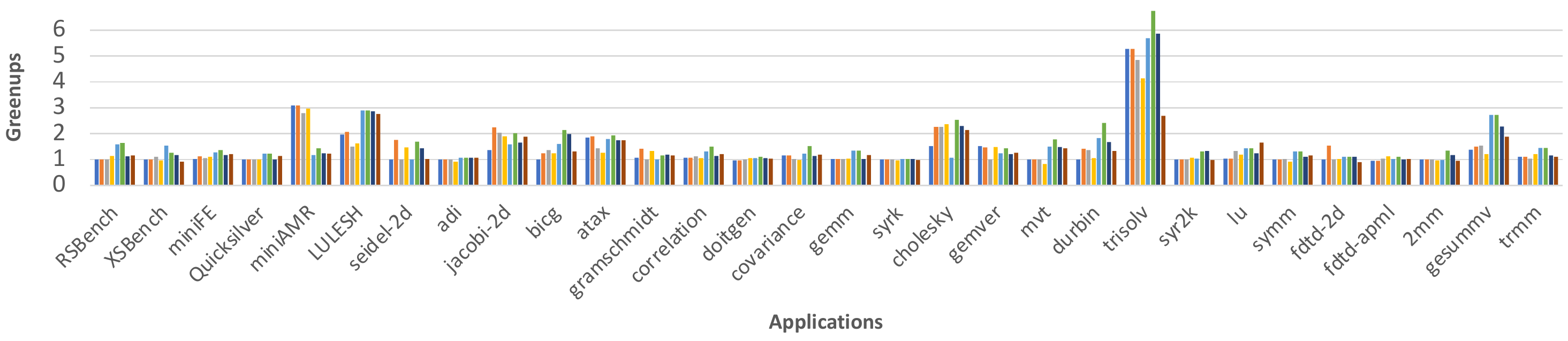} \\

\end{tabular}
\caption{Speedups/Greenups over default {\tt OpenMP} configurations at TDP. Configurations are predicted to optimize for EDP.}
\label{fig:edp_speedup_greenup}
\end{figure*}


We have also analyzed the impact on execution time performance and energy consumption individually. 
Figure~\ref{fig:edp_speedup_greenup} shows the impact of tuning for EDP on execution time for both the Skylake and Haswell systems.
Tuning for EDP leads to performance (time) improvement in $84\%$ cases, and leads to slower execution than default settings in around $16\%$ cases across both systems.
On Skylake, all slowdowns are within $20\%$ of the corresponding execution with all threads, while the geometric mean of all slowdows are within $14\%$ of the default executions.
On the Haswell system, there are fewer slowdowns, but the slowdowns are more significant with the largest slowdown within $30\%$ of the default all-threaded execution, with the geometric mean within $23\%$ of the default settings.
Overall, excluding the cases that lead to slowdowns, tuning for EDP leads to $1.16\times$ and $1.3\times$ speedups on the Haswell and Skylake.
In comparison, {\tt BLISS} and {\tt OpenTuner} leads to slowdowns in $28\%$ and $46\%$ cases respectively, with the largest slowdowns within $17\%$ and $15\%$ for {\tt BLISS} and within $30\%$ and $22\%$ for {\tt OpenTuner} on Haswell and Skylake.

We also show in Figure \ref{fig:edp_speedup_greenup} the impact of tuning for EDP on energy.
Across both systems, our approach predicts configurations that lead to reduction in energy consumption in $94\%$ cases.
In the remaining $6\%$ cases, it predicts configurations that consume more energy than the default setting.
However, the increase is minimal.
On the Haswell, there is a $3\%$ geometric mean increase in energy usage for those predictions.
On the Skylake, the corresponding number is $1\%$.
For the predictions that do lead to reduction in energy usage, there is a geometric mean greenup of $1.25\times$ and $1.42\times$ on the Haswell and Skylake respectively.
In comparison, $2\%$ of the predictions made by {\tt BLISS} lead to increase in energy consumption.
But, the overall greenups are slightly worse than the PnP Tuner ($1.24\times$ on the Haswell and $1.39\times$ on the Skylake). 
The predictions made by {\tt OpenTuner} lead to increase in energy consumption in $20\%$ cases with overall greenups at $1.25\times$ and $1.29\times$ on the Haswell and Skylake respectively.


Similar to the experiment in Section \ref{sec:pwr_const_tuning_perf_impact}, we also evaluate the effect of performance counters on EDP.
As shown in Figure \ref{fig:edp_improv}, adding performance counters to the feature set leads to improved results (predictions where the EDP is within $5\%$ of the oracle moves up to $57\%$ from $45\%$ across both systems).
Using performance counters leads to $77\%$ cases where there is improvement in execution speed (down from $84\%$).
This dichotomous behavior is the result of using a fused metric; because it is a product of both time and energy, the PnP tuner aims to tune for the best EDP.
It might lead to scenarios where the reduction in energy might compensate for the increase in time.
In this experiment, using performance counters leads to $95\%$ cases where there are improvements in energy consumption.
Overall, by using performance counters, the EDP predictions improve from $1.37\times$ to $1.52\times$ on the Haswell system, and from $1.85\times$ to $2.31\times$ on the Skylake.
This lead to overall speedups of $1.13\times$ and $1.39\times$ on the Haswell and Skylake and greenups of $1.35\times$ and $1.60\times$ on the Haswell and Skylake systems.

\section{Related Work}
This paper proposes a GNN based technique towards performance, power, and energy optimizations. 
Modifying runtime and environmental parameters has a large impact on parallel code execution. 
Several search-based tuners such as \cite{tapus2002active, ansel2014opentuner} have been proposed for these tasks.
These tuners have proposed and used several search techniques such Nelder-Mead, Torczon hillclimbers, AUC Bandit for pruning and optimizing the search space.
An alternative to search-based auto-tuning is to use machine learning (ML) based approaches. 
Several works such as \cite{alcaraz2022predicting, sanchez2020modeling, sreenivasan2019framework, thiagarajan2018bootstrapping, wood2021artemis, tehranijamsaz2022learning, dutta2022pattern, mammadli2020static} have proposed machine/deep learning based auto-tuners or tuning approaches for various parameter tuning tasks.
Recently, Bayesian optimization (BO) has been used in several works (such as \cite{sreenivasan2019framework, roy2021bliss}) for faster sampling of search spaces.
BO has gained popularity as it can be used as a black-box optimizer for an unknwown objective function.
A drawback of most of these aforementioned techniques is their need for multiple sampling executions.
Although faster than brute-force tuning, it is still a big overhead.
Our proposed static approach aims to overcome this overhead by using a \textit{static-only} approach, that does not need to execute code for tuning a fixed set of parameters.

Most of the works mentioned above do not consider power constraints in their work. 
A number of papers have focused on dynamic voltage frequency scaling (DVFS) and dynamic concurrency throttling (DCT) techniques for improving energy efficiency \cite{curtis2008prediction,li2010hybrid,de2018automatic}. 
Wang et al. in \cite{wang2015using} proposed using CPU clock modulation and concurrency throttling for improving the energy efficiency of {\tt OpenMP} loops. 
In \cite{nandamuri2014power}, Nandamuri et al. analyzed the performance and energy conumption of {\tt OpenMP} programs under various conditions using {\tt OpenMP} Runtime API. 
The work in \cite{ferreira2019performance} presented the performance and energy impact of CPU parameters on runtime systems on dense linear algebra {\tt OpenMP} kernels. 
In contrast to these works, our approach focuses on reducing energy consumption and performance improvement through power constraints.

Rountree et al. in \cite{rountree2012beyond} provided a first insight into the impacts of power capping or constraints on power and performance. Patki et al. in \cite{patki2013exploring} outlined how overprovisioning hardware and hardware enforced power bounds leads to improved performance.
To the best of our knowledge, the works in \cite{bari2016arcs,bari2019performance} are the closest to this work. Bari et al. in \cite{bari2016arcs} propose ARCS with the goal of automatically selecting best runtime configurations for {\tt OpenMP} parallel regions at specified power constraints and in \cite{bari2019performance} analyzed the impact of power constraints on performance and energy consumption on five NAS benchmarks. 
In contrast to \cite{bari2016arcs, bari2019performance}, our approach uses an AI-assisted technique based on GNNs to identify {\tt OpenMP} runtime configurations and power constraints. 
\section{Discussion}
\label{sec:limitations}
Through this study, we have outlined a unique approach to two important problems in the HPC community. 
We have proposed a mechanism of tuning {\tt OpenMP} configurations on power constrained systems. 
This is beneficial to data centers and systems working under strict power budgets.
As shown in previous sections, it is possible to considerably improve performance in such scenarios using our approach.
Additionally, we also describe a method of identifying {\tt OpenMP} configurations and power constraints that can lead to reduction in energy consumption with limited to no impact on execution time . 
To the best of our knowledge, this is the first work that aims to use GNN based techniques for these purposes. 

As with all DL techniques, training is an overhead.
Re-training a model for several target systems might be burdensome.
However, by using transfer learning techniques, we have reduced the training time on other systems by around $76\%$ on a dataset of similar size (explained in Section \ref{sec:pwr_const_tuning_exp}.
These optimizations can enable faster and easier deployment of such approaches on multiple systems.

Additionally, being a static approach, our tuner requires no sampling executions.
This is in contrast to other tuners that need several sampling runs.
Limiting the number of sampling runs, or setting a time-bound on the sampling phase to a small value leads to less than optimal results.
Moreover, our approach was able to successfully identify most edge cases.
For example, the {\tt OpenMP} region in {\tt trisolv} has the fastest execution with 1 thread in all cases.
This is an outlier.
Our approach could identify near optimal configurations in these cases as well with no executions.

Due to installation issues, we were not able to directly use the APEX framework described in \cite{bari2016arcs}. To overcome this, we used {\tt OpenTuner}, another search-based tuner as a replacement.
To contrast our work with other tuners, we present the following example.
To tune an {\tt OpenMP} region, {\tt BLISS} needs 20 sampling runs for each code region.
In case of {\tt OpenTuner}, the ``stop-after'' flag must be manipulated to allow the tuner to sample code executions.
The time bound must be increased for more complex applications, and in most cases in this paper was set to $180$ seconds and above.
A trained PnP tuner, on the other hand, needs no code executions.
\section{Conclusion}
In this work, we have outlined a twofold approach towards tuning {\tt OpenMP} configurations in power constrained systems, as well tuning both {\tt OpenMP} configurations and power constraints for execution time and energy consumption gains.
We have used GNNs to model flow-aware code graphs to model the semantic and structural features of code regions.
Our experiments show that the PnP Tuner can identify configurations that lead to improvements in execution time and energy consumption. 
In future, we aim to analyze the scalability of our approach to heterogeneous platforms and handheld devices.

\section{Acknowledgements}
This research was supported by the National Science Foundation under Grant number 2211982.
We would also like to thank the ResearchIT team (\textit{https://researchit.las.iastate.edu}) at Iowa State University for their constant support.

\bibliographystyle{IEEEtran}
\bibliography{ipdps.bib}

\end{document}